\documentclass[sigconf,screen,authorversion]{acmart}

\usepackage[normalem]{ulem}
\usepackage{soul}
\usepackage{amsmath}

\usepackage{amssymb}
\usepackage{amsfonts}
\usepackage{algorithmic}
\usepackage{graphicx}
\usepackage{enumitem}

\usepackage{hyperref}

\usepackage{pifont}

\begin{document}
\begin{sloppypar}

\title{
End-to-End Application Cloning for Distributed Cloud Microservices with Ditto}

\author{Mingyu Liang}
\email{ml2585@cornell.edu}
\affiliation{%
	\institution{Cornell University}
	\city{Ithaca}
	\state{New York}
	\country{USA}
}
\authornote{Equal contribution.}

\author{Yu Gan}
\email{yg397@cornell.edu}
\affiliation{%
	\institution{Cornell University}
	\city{Ithaca}
	\state{New York}
	\country{USA}
}
\authornotemark[1]

\author{Yueying Li}
\email{yl3469@cornell.edu}
\affiliation{%
	\institution{Cornell University}
	\city{Ithaca}
	\state{New York}
	\country{USA}
}

\author{Carlos Torres}
\email{cltorres@meta.com}
\affiliation{%
	\institution{Meta}
	\city{Menlo Park}
	\state{California}
	\country{USA}
}

\author{Abhishek Dhanotia}
\email{abhishekd@meta.com}
\affiliation{%
	\institution{Meta}
	\city{Menlo Park}
	\state{California}
	\country{USA}
}

\author{Mahesh Ketkar}
\email{mahesh.c.ketkar@intel.com}
\affiliation{%
	\institution{Intel}
	\city{Folsom}
	\state{California}
	\country{USA}
}

\author{Christina Delimitrou}
\email{delimitrou@csail.mit.edu}
\affiliation{%
	\institution{MIT}
	\city{Cambridge}
	\state{Massachusetts}
	\country{USA}
}




\date{}

\begin{abstract} 

The lack of representative, publicly-available cloud services has been a recurring problem in the architecture and systems communities. While open-source benchmarks exist, they do not capture the full complexity of cloud services. Application cloning is a promising way to address this, however, prior work is limited to CPU-/cache-centric, single-node services, operating at user level. 

We present Ditto, an automated framework for cloning end-to-end cloud applications, both monolithic and microservices, which captures I/O and network activity, as well as kernel operations, in addition to application logic. Ditto takes a hierarchical approach to application cloning, starting with capturing the dependency graph across distributed services, to recreating each tier's control/data flow, and finally generating system calls and assembly that mimics the individual applications. Ditto does not reveal the logic of the original application, facilitating publicly sharing clones of production services with hardware vendors, cloud providers, and the research community. 

We show that across a diverse set of single- and multi-tier applications, Ditto accurately captures their CPU and memory characteristics as well as their high-level performance metrics, is portable across platforms, and facilitates a wide range of system studies.

\end{abstract}

\maketitle




\section{Introduction}
\label{sec:ditto-introduction}

Cloud computing now hosts a large fraction of the world's computation, ranging from machine learning workloads to latency-critical interactive services~\cite{barroso_keynote,barrosobook,Delimitrou13, Delimitrou14,delimitrou13b,Delimitrou13d,Delimitrou14b,Delimitrou15,Delimitrou16,Delimitrou17}. Studying these applications is imperative to correctly design the systems that populate future cloud infrastructures.

There are three approaches to performing studies that require cloud applications; using real services (production or open-source)~\cite{gan:asplos:2019:microservices,usuite,cloudsuite12,Delimitrou14,googletrace,Zhang21_sosp,gan18,gan18b}, using simulation or trace replay~\cite{goyal2012cloudsim,bighouse,uqsim,stm,halo,west}, and generating synthetic services that resemble the original in behavior and characteristics~\cite{micrograd,perf_cloning,proxy_big_data,6493620,abel20nanobench,6463386}. All three approaches are subject to pitfalls. 

Using real production services is, naturally, the most representative approach. Unfortunately production services are rarely publicly available, and open-source applications, although useful, often lack the complexity and update cadence of a real cloud deployment. Studies that rely on simulation or replaying traces from a production system offer some representativeness, but are tied to the system configuration the trace was collected on, and cannot easily generalize to arbitrary studies. Finally, generating synthetic benchmarks offers a middle ground, with the synthetic application capturing critical features of the original service, but being malleable enough to adjust to different studies. Unfortunately, most prior work on synthetic benchmark cloning is limited to CPU-centric, single-tier, and user-level applications~\cite{micrograd,perf_cloning,proxy_big_data}. 

Only capturing CPU-centric microarchitectural events is not enough to reproduce the performance and resource characteristics of cloud applications, which spend a large fraction of their execution in the networking stack and OS. Moreover, prior work on synthetic application cloning mostly considers generating assembly code to mimic metrics like IPC, cache miss rate, and dependency distance, but overlooks critical higher-level performance metrics, such as average and tail latency. 

We present \textit{Ditto}, an automated application cloning framework for end-to-end cloud services, designed for both monolithic applications and microservices. Ditto is the first system to clone an application's behavior across the system stack, including the hardware, I/O, networking layers, and OS. This is critical for cloud applications which spend a large fraction of their time at kernel level and the I/O stack~\cite{kanev15}. It additionally also targets multi-tier microservices which span distributed deployments and are gaining in popularity. 

Ditto relies on the following key techniques. First, it captures the dependency graph across distributed services using distributed tracing~\cite{dapper,zipkin,jaeger,opentracing}. Then, it recreates the high-level control and data flow inside each service, and, finally, it generates system calls and user-space assembly to capture the on-CPU and off-CPU behavior. 
Ditto operates transparently to the user, with the cloning process working in an automated fashion, from obtaining a microservice deployment's dependency graph to populating each tier with appropriate assembly code and I/O operations. It generalizes across platforms, deployments, and application configurations, such as load and thread pool size, without retraining, and the synthetic applications react to changes similarly to the original ones. 

Ditto is beneficial to hardware vendors, cloud providers, and researchers. Hardware vendors can obtain synthetic versions of production applications to test new platforms, cloud providers can specify performance and/or resource specs to hardware vendors using the synthetic workloads, and researchers can use representative end-to-end cloud services without the need for production code access.

We evaluate Ditto across a set of both monolithic applications and multi-tier microservices and show that it consistently captures the low- and high-level performance metrics and resource characteristics of the original service. We also validate that synthetic applications generated with Ditto react the same way as the original workloads to changes in the input load, platform, resource allocation, and deployment configuration, including interference from external workloads and power management. Ditto is open-source software. \footnote{https://github.com/Mingyu-Liang/Ditto.}

\section{Related Work}
\label{sec:ditto-related_work}

\subsection{CPU and Cloud Benchmarking}
The architecture and system community rely heavily on software benchmarking to learn the performance characteristics of target applications. Prior studies have found that traditional CPU benchmark suites, such as SPEC~\cite{spec2017,Ganesan2010,accelerometer} and MiBench~\cite{mibench}, differ significantly from the services running in production clouds~\cite{reddi15,ueda16,softsku}. 

One of the earliest efforts towards modern cloud application benchmarking was the Cloudstone benchmark~\cite{Sobel2008CloudstoneM}, which proposed a new interaction-heavy Web 2.0 workload. CloudSuite~\cite{cloudsuite12} further composes a collection of workloads for the evaluation of scaling-out cloud services. The YCSB suite~\cite{Cooper2010} collects workloads for database systems, while SPEC Cloud~\cite{spec_cloud} utilizes a subset of workloads representing real-world use cases found on IaaS clouds. More recently, uSuite~\cite{usuite} and DeathStarBench~\cite{gan:asplos:2019:microservices} focus on benchmarking cloud microservices, given the increased popularity of this programming model. 

Instead of condensing cloud services to a pre-set group of benchmarks, Ditto enables generating arbitrary applications that resemble in features a target service. 

\subsection{Simulation and Trace Replay}
Simulation and trace replay provide another way to estimate service performance when hardware or software is inaccessible. Many microarchitectural simulators, including gem5~\cite{gem5}, Sniper~\cite{sniper} and ZSim~\cite{zsim}, can accurately simulate the CPU performance of a given binary. BigHouse~\cite{bighouse} and $\mu$qsim~\cite{uqsim} are queueing-based simulators which quickly estimate high-level performance metrics of monolithic applications and microservices. While useful when hardware is not available, these simulators still make approximations about the application behavior, and do not capture all complexities of a real system. On the other hand, RecPlay~\cite{recplay}, iDNA~\cite{idna}, PinPLay~\cite{pinplay}, Jalangi~\cite{jalangi} log the execution and memory traces, and reproduce an application's behavior for debugging and performance analysis. Unfortunately, a lot of prior work has showed that traces can leak confidential information about production services~\cite{canfora2007new,tupni,8465773}, restricting an application owner's incentive to publicly share the collected traces. West~\cite{west}, STM~\cite{stm}, HALO~\cite{halo} and Dangwal's paper~\cite{dangwal2019safer}, for example, analyze the memory access patterns of an original application, and generate synthetic memory traces. Although they can constrain the information leakage, they only target the cache and memory subsystems. Compared to trace-based techniques, Ditto generates synthetic services that clone the performance characteristics across the system stack, and can run both on real systems and microarchitectural simulators. 

\subsection{Performance Cloning and Synthetic Benchmarks}
Workload cloning is a way to generate synthetic code that mimics real-world applications. Previous studies profile architecture-independent characteristics of real applications, and generate corresponding proxy benchmarks that capture their CPU performance \\ ~\cite{bell2005improved,10.1145/1400112.1400115,6463386,proxy_big_data}. PerfProx, for example, generates miniature proxies which resemble the low-level CPU metrics of real databases~\cite{proxy_big_data}. MicroGrad~\cite{micrograd} introduces a gradient-based mechanism to generate workload clones and stress tests. NanoBench~\cite{abel20nanobench} generates microbenchmarks with certain instructions to evaluate undocumented features of x86 CPUs. In \cite{van2008dispersing}, the authors hide the functional semantics of the proprietary applications through code mutation.

However, these systems are not sufficient for performance cloning in cloud services. First, they only consider performance metrics in user space. Cloud applications spend a large fraction of their execution at kernel level ~\cite{kanev15,leverich14,gan:asplos:2019:microservices,delimitrou14,cloudsuite12,accelerometer}. Synthetic benchmarks generated with these tools focus on matching low-level performance metrics, e.g., instructions per cycle (IPC) or misses per kilo instructions (MPKI), which do not always translate to the high-level metrics cloud applications care about, like tail latency and throughput~\cite{tailatscale,beyer2016site}. Second, cloud services are often bottlenecked by off-CPU events, such as context switching or network or disk I/O, which are not captured in previous work. Finally, cloud services do not operate like independent processes, having instead client-server interfaces, which need to be captured by the cloning framework. This is even more the case for multi-tier microservices, which can have hundreds of dependent tiers, and are becoming the norm in many clouds. 


\section{Cloning Across the System Stack}
\label{sec:ditto-background}

Application cloning for cloud services is challenging due to the complexity and heterogeneity of their design, and the various platforms they can be deployed on. Different services can have entirely different bottlenecks; for example, key-value stores (KVS) require high CPU performance, high memory and network bandwidth to retrieve a large amount of data under a strict latency SLO, while databases are usually bottlenecked by disk I/O bandwidth~\cite{chen:asplos:2019:parties}. Therefore, it is important to consider the performance breakdown across the system stack to accurately clone the performance of end-to-end cloud services. 

\begin{figure}[htb]
    \centering
    \includegraphics[width=0.475\textwidth]{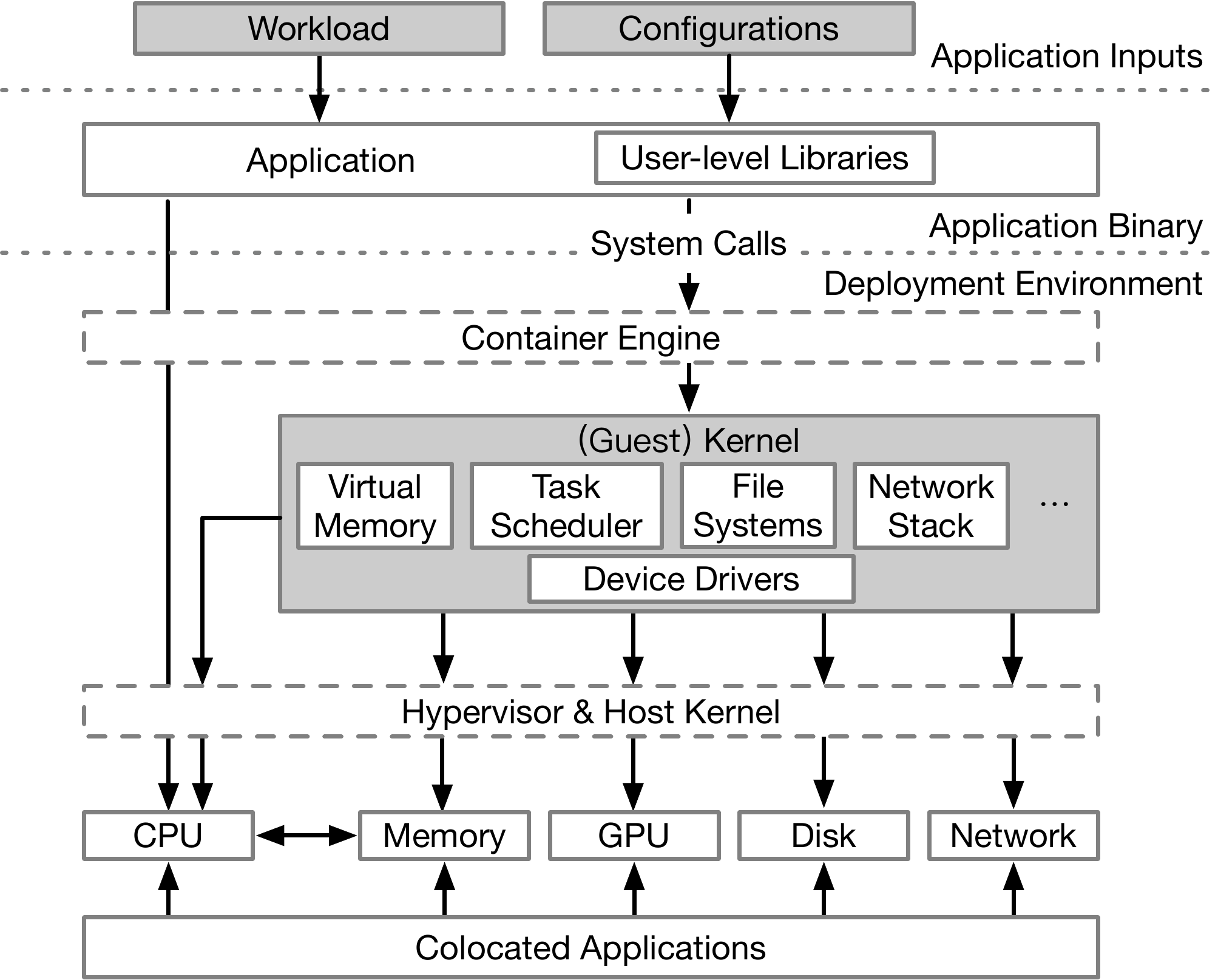}
    \caption{General system stack for cloud applications~\cite{gregg2014systems}. Dashed boxes are optional layers for virtualization. }
    \label{fig:ditto-performance_diagram}
\end{figure}

Figure~\ref{fig:ditto-performance_diagram} demonstrates an abstract view of a generic system stack for a single cloud server~\cite{gregg2014systems}. The performance of an application is determined by factors that range from the application code and inputs, to the environment it is running on, including containerization technology, the hypervisor, server platforms, and any colocated applications. We briefly describe why these factors matter below. 
\subsection{Application Inputs}
The behavior and performance of cloud applications is significantly impacted by the service configuration and input load, with the latter going through well-documented fluctuations~\cite{googletrace,meisner11,delimitrou14,delimitrou14b,accelerometer,barrosobook,211225}. 
The application's configuration, although changing less frequently than load, can substantially alter the execution flow of an application and impact performance. For instance, configuring a smaller in-memory cache for a database can cause more disk I/O accesses, significantly increasing latency. 
\subsection{Application Codebase and Binary}
The application and its linked libraries are intrinsic to its performance, regardless of the platform it is deployed on. 
Modifications in the application code can alter the control and data flow of a service, its memory access patterns, and its resource bottlenecks. This is especially true for new cloud programming frameworks, like microservices and serverless, where services are updated on a daily basis. 
\subsection{Deployment Environment}


\subsubsection{Containers and Virtual Machines (VMs)}
Cloud services are often deployed with containers and/or VMs. These add different levels of performance overheads, primarily due to the extra I/O and network layers~\cite{7095802}. Unlike prior work, Ditto faithfully clones the I/O behaviors of the cloud services, and thus, the synthetic applications generated by Ditto can be affected by virtualization the same way as the original services. 

\subsubsection{OS Kernel}
Cloud applications are especially dependent on OS performance, given that they spent a large fraction of their execution at kernel level for interrupt handling, I/O requests, memory management, task scheduling, etc.~\cite{gan:asplos:2019:microservices,dagger,kanev15,10.1145/3015146}. 
Prior work on application cloning has mostly focused on user-level application logic; for cloud services overlooking kernel operations leads to very different performance characteristics compared to the original application. 

\subsubsection{CPU-Memory Subsystem}
\begin{figure}
    \centering
    \includegraphics[width=0.475\textwidth]{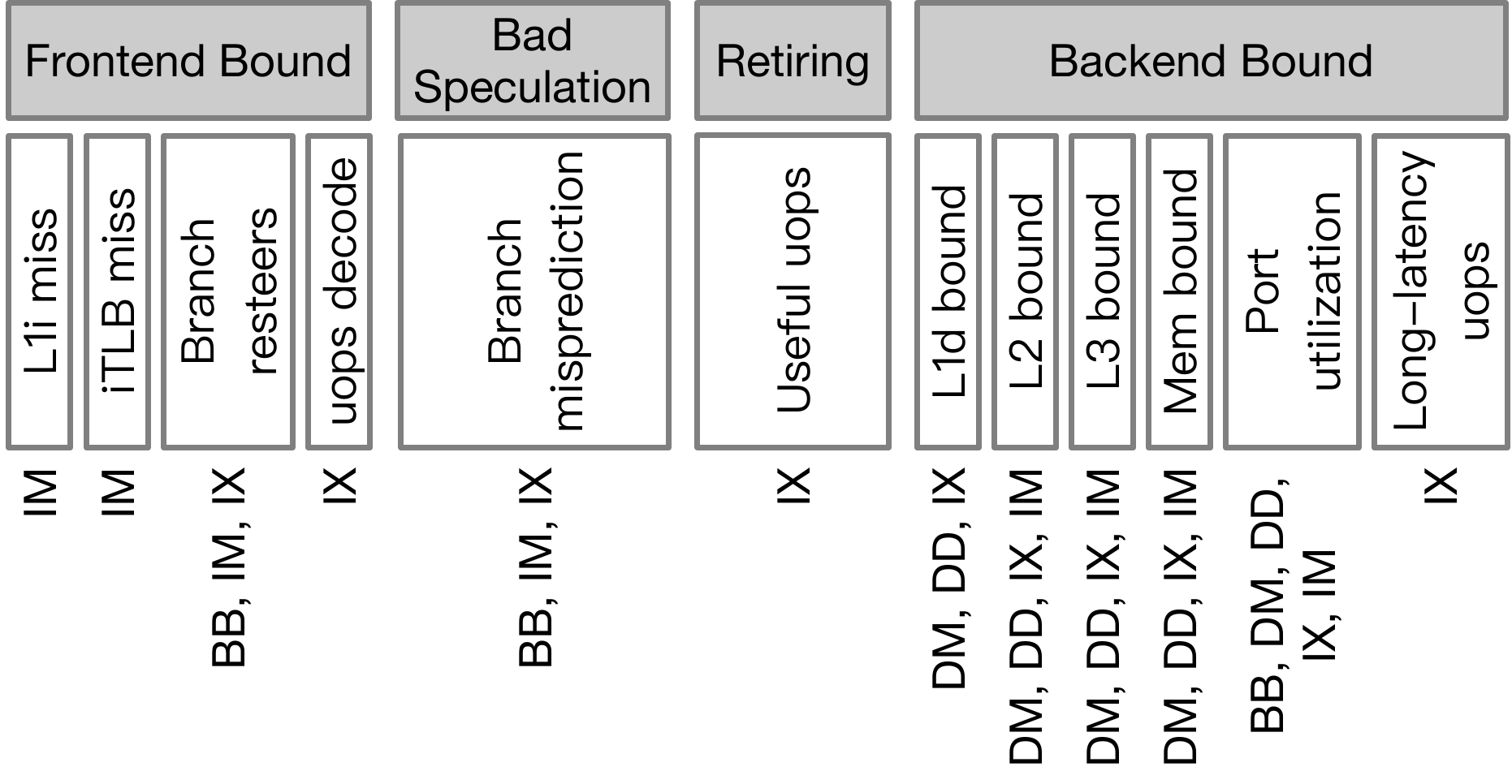}
    \caption{Top-down analysis of the CPU-memory subsystem performance~\cite{intel_topdown}. Letters at the bottom show the corresponding analysis in Ditto. \textbf{IX}: Instruction Mix. \textbf{BB}: Branch Behavior. \textbf{IM}: Instruction Memory Access Pattern. \textbf{DM}: Data Memory Access Pattern. \textbf{DD}: Data Dependency.}
    \label{fig:ditto-cpu_topdown}
\end{figure}

The CPU-memory subsystem is a dominant factor in cloud application performance, even for services that spend significant time processing network requests~\cite{dagger,kanev15,gan:asplos:2019:microservices}. We follow the top-down analysis methodology in~\cite{intel_topdown} to identify the key CPU performance metrics that impact the overall IPC and reproduce them in the synthetic applications, as shown in Figure~\ref{fig:ditto-cpu_topdown}. 
Section~\ref{sec:ditto-application_body} discusses how Ditto accounts for each of these factors during application generation. 

\subsubsection{Hardware Devices}
Services interact with hardware devices, including disks, and NICs through system calls. In cloud services specifically, peripherals can dominate performance, especially when they experience long queueing delays. We mainly consider the impact of storage and network devices in our study, as many cloud services involve I/O and network operations. Ditto can be extended to clone the behavior of other devices, such as GPUs and hardware accelerators, which we defer to future work. 
\subsection{Multi-Tenancy}


Multi-tenancy improves datacenter utilization by deploying multiple services on the same node. Applications share resources, including CPU cores, LLC, and memory, disk I/O, and network bandwidth~\cite{chen:asplos:2019:parties,lo15}. Resource contention can degrade performance, and should be accounted for in the application cloning process. 


\section{End-to-End Cloning for Cloud Services}
\label{sec:ditto-performance_cloning}

\subsection{Overview}
Ditto is an application cloning framework for cloud services; it applies to both single-tier applications and multi-tier microservices. It generates services that faithfully reproduce the performance, resource profile, and thread-level control/data flow of the original workload, decoupling representative system studies from access to the source code or the binary of production cloud services.

Ditto profiles an application at runtime and extracts key performance and resource metrics using dynamic instrumentation and runtime emulators (SystemTap~\cite{systemtap}, Valgrind~\cite{nethercote2007valgrind}, eBPF~\cite{ebpf}, Perf~\cite{perf}, VTune~\cite{vtune}, and Intel SDE~\cite{intel_sde}). Then, it generates a synthetic service which preserves the performance of the original, using an entirely distinct code sequence, to avoid revealing the implementation of the original service.

\begin{figure*}
    \centering
    \includegraphics[width=0.98\textwidth]{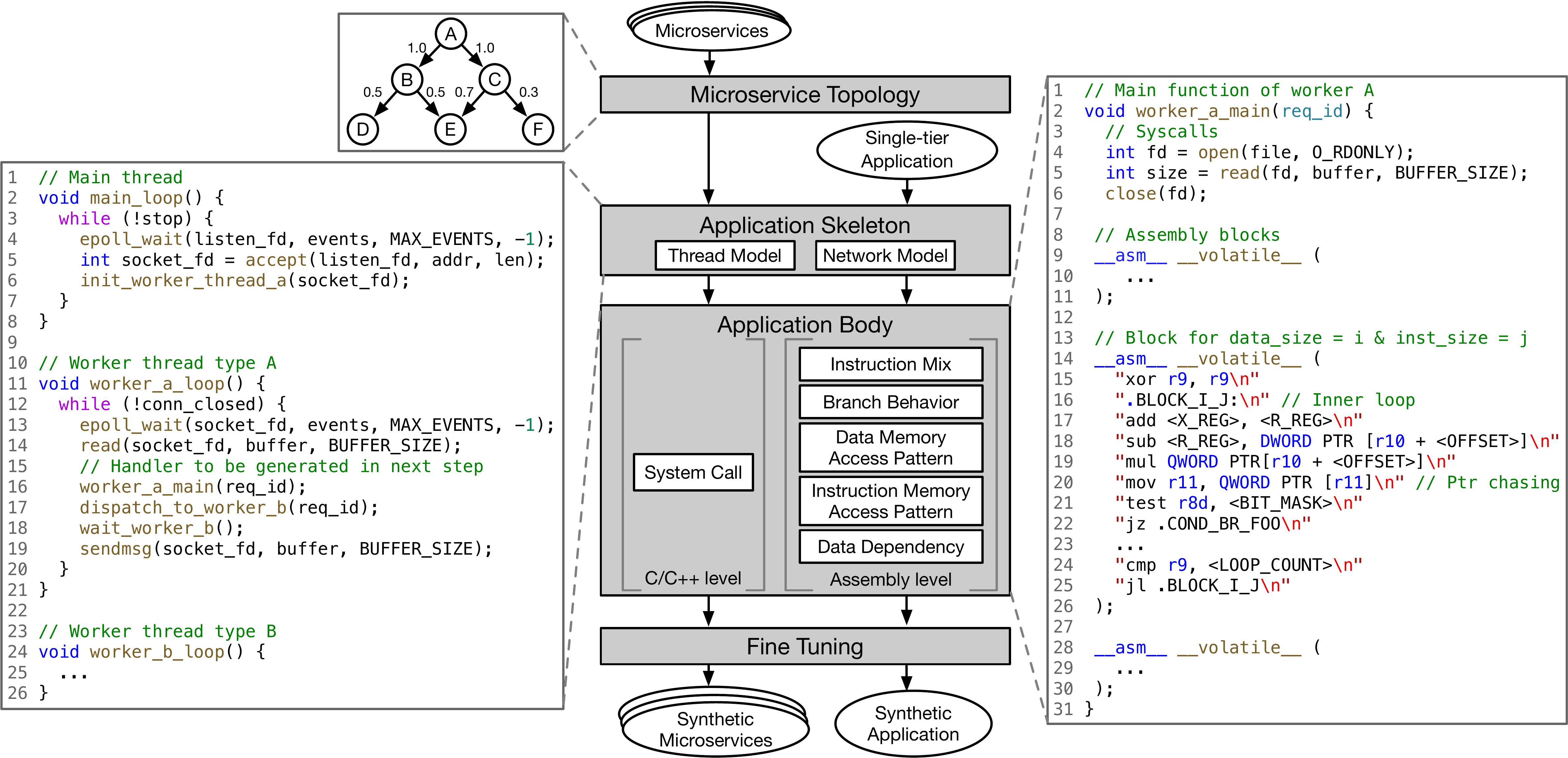}
    \caption{Overview of Ditto's synthetic benchmark generation process. }
    \label{fig:ditto-ditto_overview}
\end{figure*}

Figure~\ref{fig:ditto-ditto_overview} shows an overview of Ditto's profiling and generation process. If the target service consists of a set of microservices, Ditto first learns their Remote Procedure Call (RPC) dependency graph, using distributed tracing~\cite{dapper,zipkin,jaeger,opentracing}. This graph is then used to generate the API interfaces between the different synthetic microservices. Next, Ditto analyzes the thread and networking model, e.g., single- or multi-threaded, and synchronous or asynchronous respectively using kernel-level profiling, and builds the skeleton of each service. The application skeleton contains empty handlers which are filled with appropriate functionality in the next step. The handlers can either be triggered upon receiving requests for worker threads, or by a timer for background threads. 

To generate the synthetic application body, Ditto instruments the application binary using kernel- and user-space profilers for different subsystems. Finally, Ditto uses the deviation in performance metrics between original and synthetic application to fine tune the generator. The eventual synthetic service can serve as a performance and resource proxy for the original service. 

Ditto profiles applications in isolation to capture their characteristics alone; in Section~\ref{sec:ditto-evaluation_interference} we show that in the presence of interference, synthetic applications behave the same way as their original counterparts. 

Ditto adheres to the following design principles: 
\begin{itemize}[leftmargin=*]
\item \textbf{End-to-end system stack modeling:} Cloud services often contain a large fraction of kernel-space operations for network and disk I/O. Ditto captures the inputs, RPC dependency graph, application binary, OS kernel, CPU, memory, disk, networks, and resource interference. 
\item \textbf{Portability:} Ditto uses platform-independent features to ensure that generated services are portable across platforms without reprofiling. Synthetic applications also faithfully adjust to load and configuration changes, such as queries per second (QPS), and scaling, because of the fine-grained network and thread modeling.
\item \textbf{Abstraction:} Ditto does not disclose the implementation of the original application, only exposing the skeleton and post-processed performance characteristics to the synthetic benchmark user. It replaces the skeleton of an application with a template, refills the body with artificial instructions and their operands, and abstracts the memory access patterns away to avoid side-channel attacks. Application-specific characteristics, including user-space function calls, memory accesses, and application inputs, are also concealed. Thus, the synthetic workload can be publicly shared, without a user reverse engineering the implementation of the original service. 
\item \textbf{Automation:} Ditto automates the profiling and generation process. It entirely relies on static and dynamic profiling of the original application to generate a benchmark. Users are not required to have expertise in the implementation of a service to use the framework.
\end{itemize}

\subsection{Microservice Topology}
A topology of microservices is a directed acyclic graph (DAG), where the nodes are microservices and the edges indicate the dataflow between dependent tiers~\cite{gan:asplos:2019:microservices,gan:asplos:2019:seer,sage, alibaba_trace}. Ditto leverages the distributed trace frameworks present in most production deployments to collect traces of end-to-end requests. The performance overhead is negligible if the traces are sampled properly~\cite{dapper,zipkin,jaeger}. It then automatically extracts the dependency graph between microservices and uses it as input to the skeleton generator.

\subsection{Application Skeleton}
\label{sec:ditto-skeleton}

We define the application skeleton as the network and thread models of an application, which determine how it handles remote service communication, and how tasks are assigned to different threads, respectively. The application skeleton is a critical design choice for cloud services facing tight latency constraints~\cite{utune, pariag2007comparing, fan2015performance, wang2017study}, as it directly impacts their performance and scalability. 

\subsubsection{Network Model}
\label{sec:ditto-technique_network_model}

The network model describes how an application communicates with other services, acting as a client, server, or both. When acting as a client, a service can use synchronous or asynchronous communication. In synchronous models, threads block on network I/O (e.g., \texttt{send()}, \texttt{write()}) to await responses. Asynchronous models are typically event-based with responses handled by specific threads via callback functions. They are more complicated, as they involve additional synchronization and state machine transitions. In return, they avoid long queueing delays by allowing threads to process new requests and offer better performance ~\cite{utune}. 

On the server side, there are three common options for the network model: blocking, non-blocking, and I/O multiplexing~\cite{stevens1990unix}. In all three models, threads await requests through system calls (e.g., \texttt{recv()}, \texttt{read()}, \texttt{epoll()}). In contrast to the other two models, the non-blocking model needs to periodically call the I/O interfaces to look for new requests, which can waste CPU time at low loads. In both blocking and I/O multiplexing models, threads block on system calls, although I/O multiplexing allows monitoring multiple sockets via a single system call (e.g., \texttt{select()} or \texttt{epoll()}). I/O multiplexing is the most commonly-used in services like Memcached, Redis, and NGINX, since they support many concurrent connections, and I/O multiplexing reduces the required threads. 


Ditto uses SystemTap~\cite{systemtap} to profile the network model by probing kernel-space functions and data structures. It acquires key attributes of sockets, and monitors network-related system calls, gathering the distribution of their types, arguments, and call frequency. Ditto then chooses one out of several network models that combine the different design choices described above, with socket options and network message parameters set based on profiling. 


\subsubsection{Thread Model}

Cloud services rely on multithreading for asynchronous networking, disk I/O, and parallel processing~\cite{reddi15}. The thread model describes how tasks are scheduled to and handled by various threads. Ditto uses SystemTap to profile the functionality, lifecycle, and trigger points of threads by experimenting with different connections, QPS, and execution times. First, it combines network and user-space call stack analysis to cluster threads with similar functionalities. We build a call graph for each thread, use tree-edit distance~\cite{bille2005survey} to measure the similarity between threads, and cluster threads with similar call graphs using agglomerative clustering~\cite{murtagh2012algorithms}, since the number of clusters is unknown in advance. Second, we categorize each thread cluster into short- and long-lived threads by probing \texttt{clone()} and context switches. Short-lived threads are usually spawned and terminated frequently, while long-lived threads are spawned at initialization, waiting for tasks to arrive. Finally, thread functions can be triggered by both kernel- and user-space events, including reads and writes to sockets, timers, signals, user-space locks, and condition variables. We monitor event notification functions in kernel space and common user-level libraries, such as \texttt{libpthread} and \texttt{libc++}, and analyze the relationship between them and thread spawning or wakeup to identify trigger points. 



\subsection{Application Body}
\label{sec:ditto-application_body}
The application body corresponds to the workload-specific work, consisting of kernel-space functions, via system calls and user-level functions. While assembly-level profiling for kernel-space functions is unnecessary, since they can be cloned by imitating the system calls themselves, it is critical to clone user-space functions at assembly level to capture the low-level usage of CPU resources. 

Application performance is also significantly impacted by factors like instruction mix, and memory (data and instruction) access patterns, branch behavior, and data dependencies. Ditto uses these platform-independent features to ensure that the generated synthetic applications can be ported to other platforms without reprofiling. 

\subsubsection{System Calls}
Applications use system calls to perform privileged operations in the OS kernel. Besides network handling and spawning new threads, cloud applications can make system calls to access file descriptors, allocate memory space, or synchronize on shared memory. Capturing the system call characteristics is critical to clone the kernel-level CPU and un-core metrics. Prior performance cloning studies either do not profile system call characteristics~\cite{perf_cloning,micrograd}, or only profile the total number of kernel-level instructions~\cite{proxy_big_data}. To accurately capture kernel-level characteristics, Ditto profiles the distribution of system calls, including their counts and arguments with SystemTap. For example, MongoDB calls \texttt{pread()} to read a database file from disk. During system call profiling, Ditto captures the flags of \texttt{fd} and the distribution of \texttt{count} and \texttt{offset}, to accurately clone key metrics, such as disk latency, utilization, and page cache miss rates.

\subsubsection{Instruction Mix}
The instruction mix in Ditto captures the distribution of x86 assembly instructions at runtime in the original service, and reproduces it faithfully in the synthetic benchmark. 
Previous studies categorize x86 assembly instructions into integer arithmetic, integer multiplication, integer division, floating-point operations, SIMD operations, loads, stores, and control instructions~\cite{perf_cloning,proxy_big_data,micrograd}. They then generate the synthetic benchmark using a representative instruction from each category.

However, this categorization is too coarse-grained and does not capture the characteristics of modern CPU microarchitectures. The x86 ISA, for instance, contains assembly instructions with different uops, port usages, and execution cycles. For example, the \texttt{CRC32 (r64, r64)} instruction, which implements the checksum function, takes three cycles and can only be executed via port 1 on Skylake CPUs, while other integer arithmetic instructions usually take one cycle on any of the ports 0, 1, 5, and 6~\cite{uops.info,guide2011intel}. 
Instructions with \texttt{REP/REPZ/REPNZ} (repeat string operations) or \texttt{LOCK} prefixes can take tens of cycles or more, depending on the repeat count, or the cache/RAM configuration~\cite{fog2011instruction}. 

Ditto uses Intel SDE~\cite{intel_sde} to collect the dynamic count of each x86 instruction using Intel x86 Encoder Decoder (XED) Iforms~\cite{x86_xed}. It then clusters x86 assembly instructions by functionality (data movement, arithmetic/logic, control-flow, lock-prefixed, and repeat string operations), operands (general-purpose registers, x87 floating-point registers, XMM registers, and memory), and ALU usage~\cite{uops.info} using hierarchical clustering, so that each cluster has similar hardware resource requirements. Ditto also profiles the average number of dynamic instructions per request, and the repeat counts 
of each \texttt{REP}-prefixed instruction. During the generation phase, Ditto randomly samples the next instruction from the instruction mix distribution. Registers and memory addresses are assigned after data memory access profiling (Section~\ref{sec:ditto-dmem_access_pattern}). 

\subsubsection{Branch Behavior}
\label{sec:ditto-branch_prediction}
Branch prediction accuracy, which is determined by both the branch behavior of the application and the branch predictors, is critical in modern out-of-order CPUs~\cite{9042108,guide2011intel}. Prior studies observe that branch taken ratios and transition rates (frequency a branch switches between taken and not-taken) impact the branch prediction accuracy and misprediction penalty~\cite{fog_microarchitecture,824354}. Branches with extremely high taken or not-taken ratios, even if their patterns are completely random, have fewer mispredictions, since the majority of executions are in one direction. Similarly, branches with low transition rates are easier to predict. 
We also find that instruction locality and the number of static branch instructions significantly contribute to the branch prediction accuracy, especially for applications with large binaries. 


% ...


Based on these observations, Ditto profiles the distribution of branch taken/not-taken rates and transition rates across all conditional branch instructions, and together with the instruction memory access pattern analysis it accurately clones the branch misprediction behavior of the target application. We quantize the taken/not-taken rates and transition rates in log scale, from $2^{-1}$ to $2^{-10}$. During the generation phase, Ditto samples a taken/not-taken rate and transition rate from the profiled distribution for each conditional branch instruction. 

Lines 21-22 in the right code snippet in Figure~\ref{fig:ditto-ditto_overview} show how Ditto generates conditional branch instructions with profiled taken/not-taken rates and transition rates. \texttt{<BIT\_MASK>} is a binary mask pre-computed during the generation phase, which contains $M$ ones in the highest bits and $N$ zeros in the lowest bits. $2^{-M}$ is the taken/not-taken rate, and $2^{-N}$ is the transition rate. The \texttt{ZF} flag, which determines the branch direction of \texttt{jz} or \texttt{jnz}, will change periodically according to the bitmask in the \texttt{test} instruction.

\subsubsection{Data Memory Access Pattern}
\label{sec:ditto-dmem_access_pattern}
The memory access pattern is a dominant characteristic of an application, as it impacts the backend of the CPU and memory subsystem. Since operands in arithmetic instructions in synthetic benchmarks are randomly generated, they cannot calculate meaningful memory addresses at runtime. Thus, memory addresses or offsets need to be pre-calculated in the generation phase and hard-coded in the synthetic application binaries. Previous studies~\cite{west,stm,halo} capture memory access patterns using the stack distance, reuse distance, and stride pattern profiles. However, they need 10 to 20 million memory traces to accurately represent target memory access patterns because of the sparsity of the memory address space, and the multimodality in memory accesses~\cite{pmlr-v80-hashemi18a}. Preserving the original access patterns requires millions of hard-coded memory instructions, which significantly interferes with other performance characteristics.
Moreover, directly replicating the target memory access pattern introduces security concerns, since previous studies showed that memory access patterns reveal confidential information about the service~\cite{8465773,goldreich1996software,8141932}. 

\begin{figure}[htb]
     \vspace{-0.08in}
    \centering
    \includegraphics[width=0.43\textwidth]{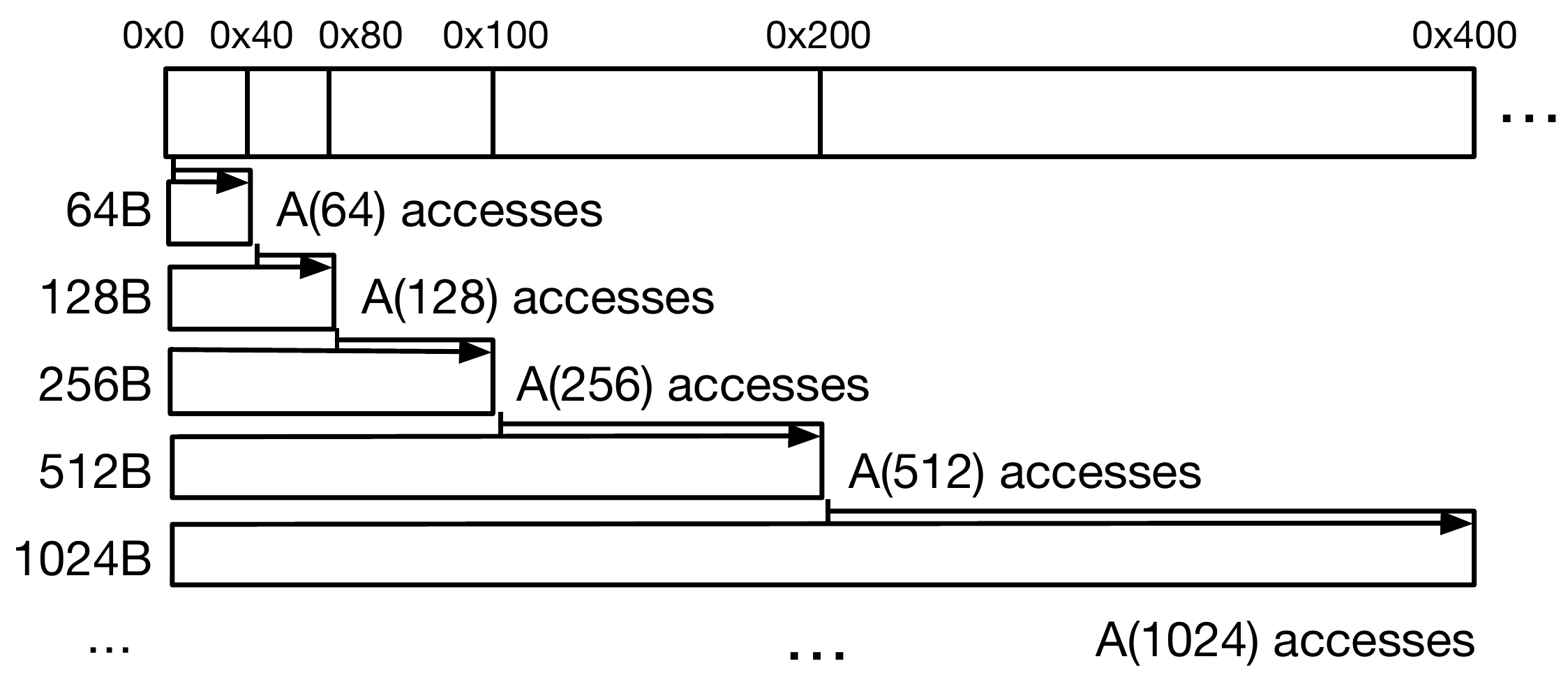}
    \caption{Working-set-based data memory access generation. Except for the 64-byte working set, the memory accesses of $2^i$-byte working set start at address $2^{i-1}$ and loop iteratively within the working set. }
    \label{fig:ditto-memory_access_example}
     \vspace{-0.18in}
\end{figure}

Instead, Ditto uses profiling of the memory working set to synthesize appropriate data memory access patterns without incurring high instruction misses or leaking application context. We construct a sequence of memory accesses for working sets with different sizes, from 64 bytes (one cache line) to the maximum memory size allocated to the target application, increasing by a factor of two. Each memory access only reads or writes the first data in a cache line to ensure that a new cache line is loaded, assuming the most common write-allocate policy. We use Valgrind~\cite{nethercote2007valgrind} to compute the distribution of memory accesses with different working set sizes, which can be efficiently simulated as ``cache hits'' for different ``cache sizes''. Each ``cache size'' only needs to be simulated once during profiling. We calculate the number of memory accesses in a working set of $2^i$ bytes as follows:

\begin{equation}
A_d(2^i) = 
\begin{cases}
H_d(2^i) & \text{if $2^i=64$ bytes} \\
H_d(2^i) - H_d(2^{i-1}) & \text{otherwise}    
\end{cases}
,
\end{equation}

where $A_d(2^i)$ is the number of memory accesses for a working set of $2^i$ bytes in generated code, and $H_d(2^i)$ is the number of cache hits in a $2^i$-byte cache in the original application. The synthetic working set-based memory access pattern is illustrated in Figure~\ref{fig:ditto-memory_access_example}, with the number of memory accesses for each working set equal that of the profiled distribution. Since the memory accesses are limited to the working set size, it is guaranteed that $A_d(2^i)$ accesses will contribute to $A_d(2^i)$ hits when cache size $\ge 2^i$ bytes. Assuming a least-recently-used (LRU) cache replacement policy or its pseudo-LRU variant, commonly used in recent Intel processors~\cite{abel20nanobench,cachequery}, since we iterate through cache lines in a working set sequentially, there must be previous memory accesses which evict this cache line when cache size $< 2^i$ bytes. Therefore, every memory access of a $2^i$-byte working set ends up with a miss when cache size $< 2^i$ bytes. The statement is true for any memory hierarchy and cache inclusion policy because of the sequential access pattern within each working set. Therefore, even if applications are profiled with a single-level cache, the results can be applied to any number of cache levels and inclusion policies. Applications are profiled with an 8-way cache for working sets < 1MB and a 16-way cache for working sets $\geq$ 1MB, which are close to the typical values of modern CPUs. There is an average 1.9\% error in the cache miss rate when cache associativity changes across all examined applications. We allocate an array for memory accesses in the heap when the synthetic application is initialized, and store the base address in a register (for example, \texttt{r10}). Ditto generates the address offsets for each memory instruction, which can access \texttt{[r10 + <OFFSET>]} at runtime. 

Coherence misses also contribute to cache miss rates in multi-threaded applications. Coherence misses happen when cache lines containing shared data are invalidated by another core. To accurately clone cache behavior with multi-threading, we use Intel SDE to profile the ratio between private data accesses and shared data accesses, and generate memory accesses accordingly. 

Modern CPUs implement hardware prefetching mechanisms to improve cache performance. Hardware prefetchers detect load instructions with regular strides, sequences of consecutive cache line accesses and adjacent cache line accesses, to load data into caches before they are needed~\cite{intelOptManual}. To clone the performance impact of cache prefetching, we calculate the ratio of regular to irregular memory access patterns from the runtime memory trace and use this ratio to control the number of regular memory access sequences in the synthetic applications. 

\vspace{-0.09in}
\subsubsection{Instruction Memory Access Pattern}
Instruction memory access patterns significantly impact CPU frontend and backend performance, as they determine the L1i, L2, L3 cache misses and branch mispredictions. Replicating the original application's instruction memory access pattern is not possible with a synthetic benchmark because the execution flow is usually controlled by the computation's output at runtime.

Therefore, Ditto synthesizes instruction memory access patterns with a similar approach to that of Section~\ref{sec:ditto-dmem_access_pattern}. We profile the i-cache hits of the original application with different i-cache sizes using Valgrind. Then, we calculate the distribution of dynamic executions in an instruction memory working set of $2^j$ bytes as follows, assuming the cache line size is 64 bytes, and the average instruction size is 4 bytes:
\begin{equation}
\label{eq:ditto-icache}
E_i(2^j) = 
\begin{cases}

16 * \big[H_i(2^j) - H_i(2^{j-1})\big] & \text{if $2^j>64$ bytes} \\
H_i(2^N) - \sum_{j=9}^{2^N}{E_i(2^j)} & \text{if $2^j=64$ bytes} 
\end{cases}
,
\vspace{-0.06in}
\end{equation}

where $E_i(2^j)$ is the number of instruction executions with a working set of $2^j$ bytes in the synthetic code, $2^N$ is the max instruction working set size, $H_i(2^j)$ is the number of i-cache hits on a $2^j$-byte i-cache in the original application, and the number of instructions in a cache line is 16 (64B cacheline / 4B inst size). After profiling the distribution of i-cache accesses with different instruction working sets, Ditto generates static assembly instruction blocks, shown in lines 14-26 in the right code snippet of Fig.~\ref{fig:ditto-ditto_overview}.

The number of instructions per block matches the instruction working set size, and the loop iteration number is determined by the distribution. 

\vspace{-0.09in}
\subsubsection{Data Dependencies}
Data dependencies are another inherent characteristic of an application that impact performance. Data dependencies can flow through registers or memory locations, limiting the number of simultaneous instructions issued to an execution unit (instruction-level parallelism, or ILP), and the number of outstanding memory requests (memory-level parallelism, or MLP)~\cite{hennessy06}. 

Ditto uses the distribution of data dependency distances to quantify data flows through registers. We measure the read after write (RAW), write after read (WAR), and write after write (WAW) data dependency distance from the dynamic control flow graph (DCFG) generated using Intel SDE~\cite{intel_dcfg}. The dependency distance is quantized into 11 bins, increasing exponentially from 1 to 1024, since a larger dependency distance does not impact the ILP, due to the limited size of the reorder buffer. When generating the synthetic code, we reserve several registers for recording the loop counters and data memory addresses, and use the rest of general-purpose and SIMD registers to clone the data dependency characteristics. To assign registers for each instruction, Ditto samples a (RAW, WAR, WAW) distance tuple from the profiled distributions, and chooses an available register with the closest distance values. Data dependencies through registers can also impact MLP if the register values determine memory locations. Such behavior cannot be captured since the synthetic application never writes to a reserved register with the memory base address. To address this, we replace a fraction of memory reads with pointer chasing reads (\texttt{mov r11, QWORD PTR [r11]}); determined by the MLP measured with Perf.

Data dependencies through memory locations are much more difficult to profile with DCFG since memory addresses are often calculated at runtime. However, they are partially determined by data access patterns that Ditto already profiles (Section~\ref{sec:ditto-dmem_access_pattern}). A program with a shorter memory dependency distance can be modeled with a smaller working set because the probability that the cache line is evicted by other instructions in between is lower. 


\vspace{-0.1in}
\subsection{Fine Tuning}
Finally, Ditto implements fine tuning to calibrate the output of previous steps, due to inaccuracies introduced by the instrumentation tools. For example, application body profiling does not consider the interaction between user-space and kernel-space functions and the correlation between the application skeleton and body; thus the actual d-cache and i-cache miss rates are often higher than the profiled results. Ditto iteratively runs the synthetic application on a specific platform, computes the errors between target and synthetic service, adjusts the inputs to the generator accordingly, and regenerates the synthetic application. Although there are many knobs to tune, most of them are orthogonal with each other. We have characterized the correlation across knobs, and derived the small groups of parameters that need to be jointly tuned (e.g., branch taken/transition rate and i-cache pattern because they all influence branch prediction). Since relationships between knobs and performance are mostly linear, we use a feedback-based heuristic to tune knobs within a group. Fine tuning uses performance counters for calibration. It usually takes within ten iterations to reach over 95\% accuracy, incurring low overhead since each iteration only takes a couple tens of seconds. Since Ditto captures performance characteristics well with platform-independent data, this fine tuning does not compromise the generality of the synthetic service, as shown in Section~\ref{sec:ditto-validation_varying_platforms}.

\vspace{-0.1in}
\section{Implementation}
\label{sec:ditto-implementation}
Ditto implements several analyzers and code generators to capture the microservice topology, application skeleton, and application body. If the target service is a graph of microservices, the microservice topology analyzer leverages distributed tracing systems, like Jaeger~\cite{jaeger}, to obtain RPC call graphs and call statistics. For both microservices and single-tier applications, the application skeleton analyzer then deploys SystemTap to profile network- and thread-related functions and data structures in kernel space, and identify the network and thread models used.  

The skeleton generator creates a synthetic application skeleton using either a TCP- or RPC-based network interface, leaving the body of each thread's handler to the application body generator. The latter runs SystemTap to profile system calls, and uses Intel SDE and Valgrind to capture the platform-independent features of binaries, such as instruction mix and working set size distribution, etc. The generator creates handlers according to these features using POSIX APIs in libc and inline assembly in C code. The assembly code contains tens to hundreds of instruction blocks looping iteratively with different instruction and working set sizes. Finally the fine tuner runs the synthetic application, collects performance data from Perf, eBPF and VTune on the deviation between original and synthetic workloads, and calibrates the input data for the application body generator accordingly. 


Ditto is implemented primarily in Python and C in about 16,000 lines of code. It supports C/C++ applications, the Apache Thrift~\cite{thrift} and gRPC~\cite{grpc} RPC frameworks, and x86 ISAs, which are commonly used in cloud environments. It can be extended to more languages, frameworks, and ISAs, by leveraging compatible profiling tools. Ditto can generate applications that run on a single machine or containerized microservices that run on multiple nodes, using Docker Swarm or Kubernetes. The runtime profilers and emulators, including SystemTap, Intel SDE, and Valgrind, can introduce overheads to the original application during profiling. This overhead only occurs once, and does not affect the accuracy of the platform-independent features collected during profiling. 

To generate a clone, cloud providers only need to specify a representative input for their service. Ditto automatically instruments the application at runtime, collecting profiling statistics and feeding them to the code generator, followed by the fine-tuning process. Ditto does not require reprofiling if the input change does not affect the application body, such as changes in QPS or number of connections. Inevitably, if a new input exercises an entirely new code path or memory access pattern, this will need to be profiled to create a new clone. The synthesized binaries can run directly on hardware, execution-driven simulators like gem5~\cite{gem5} and ZSim~\cite{zsim}, or their traces can be fed to trace-driven simulators like Ramulator \cite{kim2015ramulator}.



\section{Evaluation}
\label{sec:ditto-evaluation}

\subsection{Methodology}

\subsubsection{Platforms}
We validate Ditto on a heterogeneous cluster, with three types of servers, whose specs are in Table~\ref{tab:ditto-platform_spec}. All servers run x86 ISA, but differ in the CPU and memory architectures, and their storage and network. 

\vspace{-0.1in}
\begin{table}[htb]
    \centering
    \caption{Server platform specifications.}
    \vspace{-0.1in}
    \resizebox{0.95\columnwidth}{!}{%
    \begin{tabular}{c|ccc}
    \hline\hline
                         & Platform A & Platform B & Platform C \\ 
        \hline
         CPU model       & Gold 6152 & E5-2660 v3 & E3-1240 v5 \\
         Base Frequency   & 2.10GHz & 2.60GHz & 3.50GHz \\
         CPU cores       & 22 & 10 & 4 \\
         CPU family      & Skylake & Haswell & Skylake \\
         Sockets         & 2 & 2 & 1 \\
         L1i/L1d         & 32KB/32KB & 32KB/32KB & 32KB/32KB \\
         L2              & 1MB & 256KB & 256KB \\
         LLC             & 30.25MB & 25MB & 8MB \\
         RAM             & 192GB@2666 & 128GB@2400 & 32GB@2133  \\
         Disk            & 1TB SSD & 2TB HDD & 1TB HDD \\
         Network         & 10Gbe   & 1Gbe    & 1Gbe \\
     \hline\hline
    \end{tabular}
    }
    \label{tab:ditto-platform_spec}
    \vspace{-0.2in}
\end{table}

\begin{figure*}
    \centering
    \includegraphics[width=0.95\textwidth]{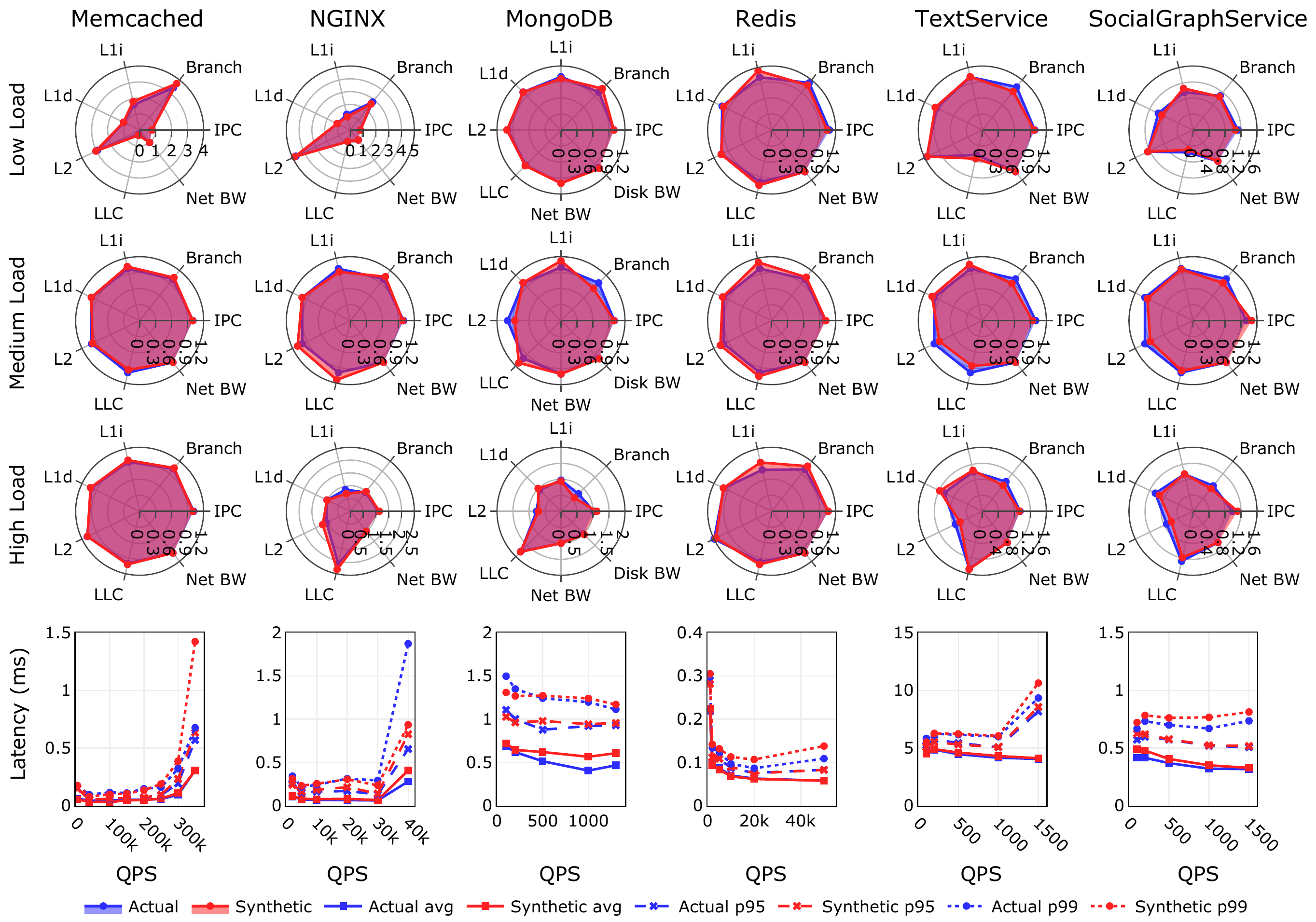}
    \vspace{-0.1in}
    \caption{CPU performance metrics (IPC, branch mispredictions, L1i, L1d, L2 and LLC miss rates), network bandwidth, disk bandwidth (MongoDB only) and service latency under varying load across six services. CPU metrics are normalized to each original application's metrics under medium load. Network and disk bandwidth are, by exception, normalized to each original application's bandwidth under current load, because their magnitudes change significantly, and would obscure the figure's shape.}
    \label{fig:ditto-cpu_qps}
    \vspace{-0.12in}
\end{figure*}

\subsubsection{Applications and Workload Generators}
\begin{itemize}[leftmargin=*]
\item \textbf{Memcached:} Memcached~\cite{memcached} is a distributed low-latency, key-value store for in-memory caching. We build Memcached 1.6.9 from source, deployed with four worker threads, and load it with 10K items, each with a 30B key and a 4KB value. It is driven by an open-loop version of the mutated workload generator~\cite{mutated}.
\item \textbf{NGINX:} NGINX~\cite{nginx} is a high-performance web server and is the most commonly-deployed technology in Docker~\cite{docker_2018_container_survey}. We build NGINX 1.20.0 from source and configure it with one worker process. For NGINX, we use tcpkali~\cite{tcpkali} to generate HTTP requests. 
\item \textbf{MongoDB:} MongoDB~\cite{mongodb} is an open-sourced cross-platform NoSQL database. We use MongoDB 4.4.4 and set up a dataset of 40GB with one million records. To load MongoDB, we use YCSB~\cite{Cooper2010} with all read operations, following a uniform distribution.
\item \textbf{Redis:} Redis~\cite{redis} is a fast, single-threaded, in-memory data store used as a database, cache, and message broker. We build Redis 6.2.6 from source, disable its persistent storage, and load a dataset with 100K records. We use YCSB as the load generator. 
\item \textbf{Social Network:} Social Network is a microservice topology from DeathStarBench~\cite{gan:asplos:2019:microservices}, consisting of 20+ individual services. We compose its social graph with the socfb-Reed98 Facebook dataset~\cite{reed98}, which contains 962 users and 18.8K follow relationships. We also modify the wrk2~\cite{wrk2} workload generator to open-loop and use it as the client. The Social Network is deployed with one replica per microservice, both locally and on a cluster using Docker containers.
\end{itemize}

For all synthetic applications, we use the same load generator as the original application, sending dummy requests with the same traffic distribution. The number of threads of MongoDB and Social Network microservices changes dynamically with the number of concurrent connections, up to a few tens under our load settings.

\begin{figure}
    \centering
    \includegraphics[width=0.43\textwidth]{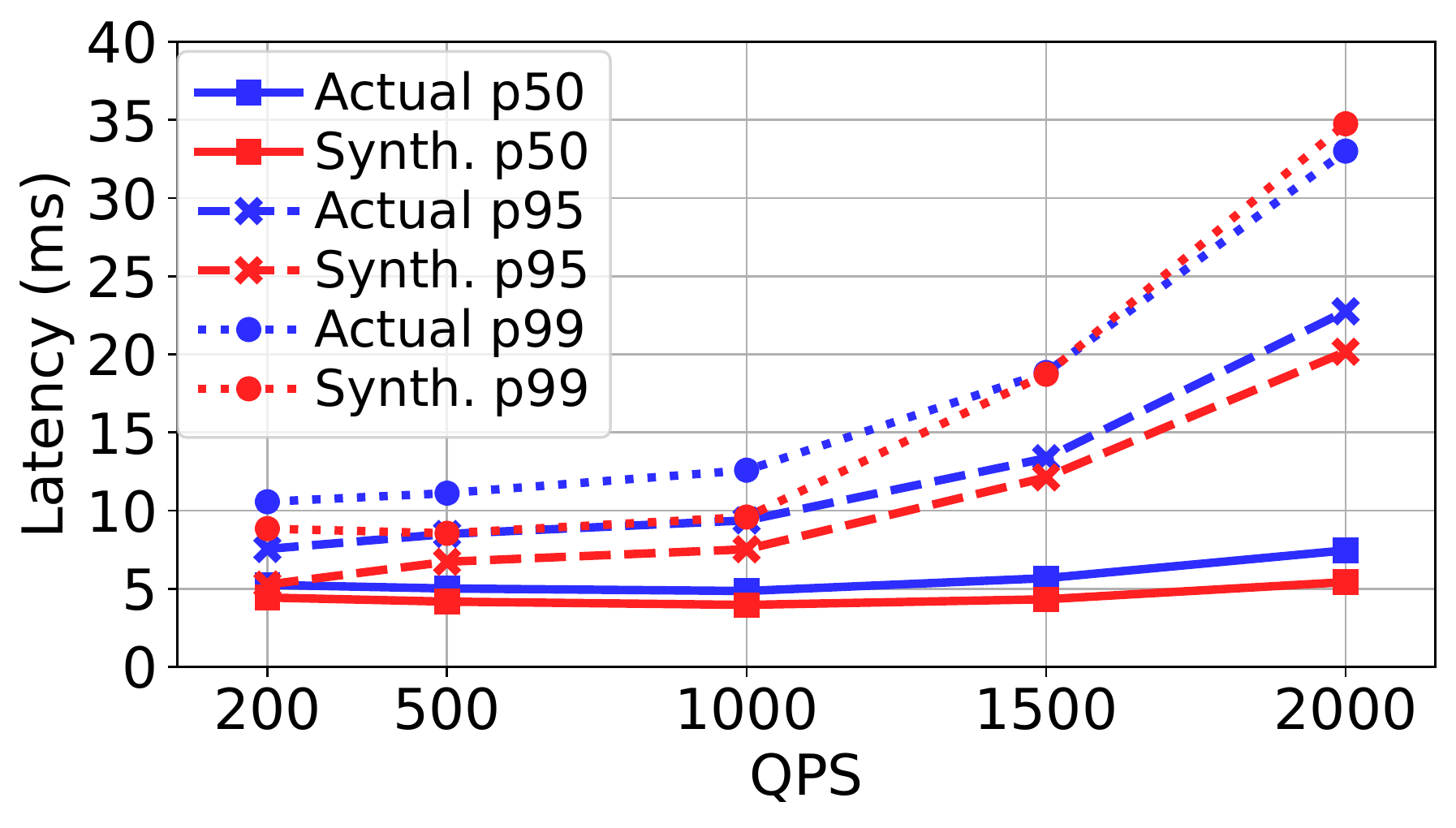}
     \vspace{-0.12in}
     \caption{End-to-end latency for the Social Network.}
    \label{fig:ditto-e2e_latency}
     \vspace{-0.25in}
\end{figure}

\begin{figure*}[htb]
    \centering
    \includegraphics[width=0.95\textwidth]{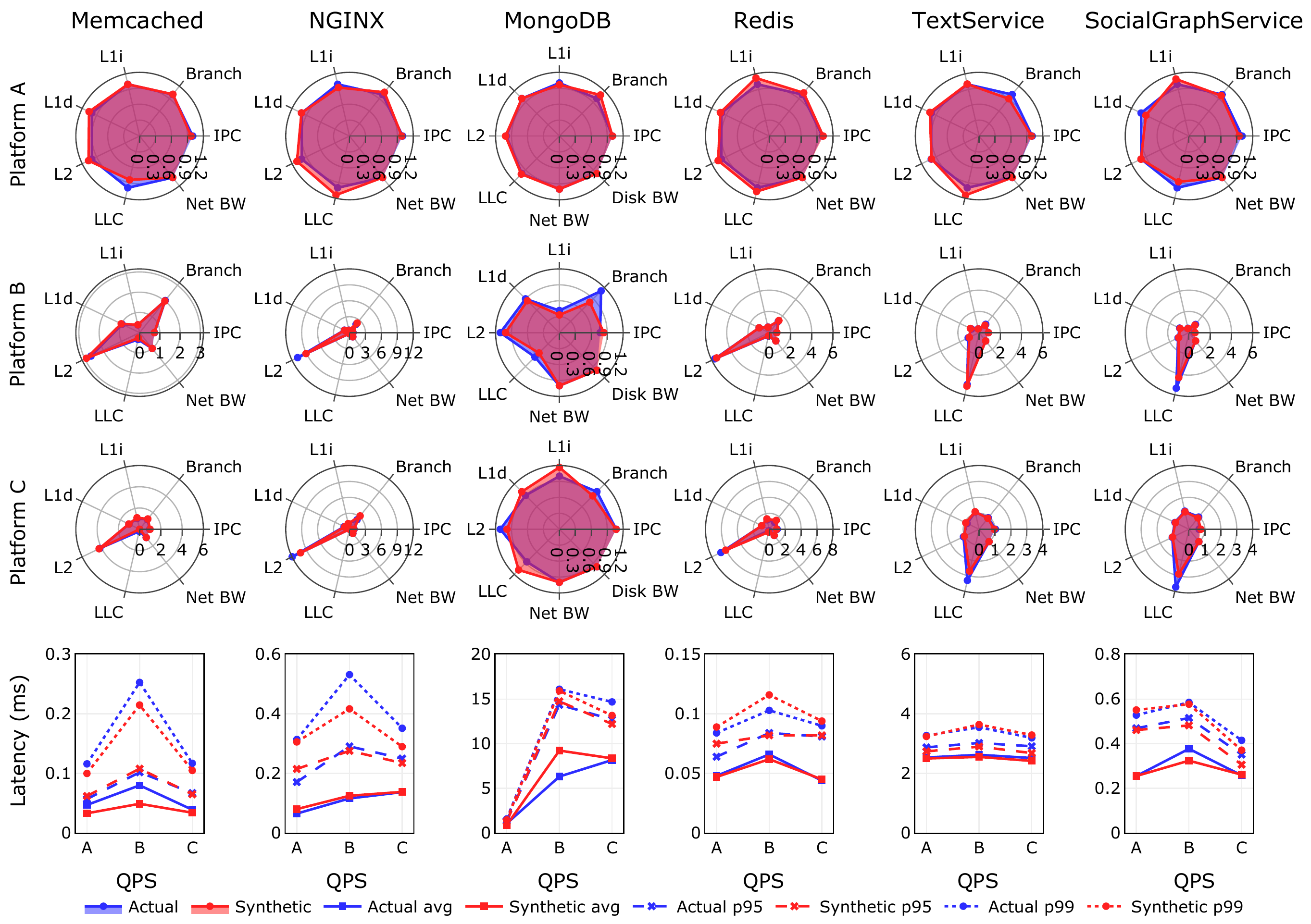}
    \caption{CPU metrics (IPC, branch misprediction, L1i, L1d, L2 and LLC misses), network BW, disk BW (MongoDB only) and latencies across platforms. CPU metrics are normalized to each original service on Platform A.}
    \label{fig:ditto-cpu_platform}
    \vspace{-0.05in}
\end{figure*}




\subsection{Validation}
\label{sec:ditto-validation}

\subsubsection{Validation on Varying Loads}
Figure~\ref{fig:ditto-cpu_qps} shows CPU, network and disk performance metrics, and latency for six applications under different QPS in platform A. In addition to the four single-tier applications, we also show resource characteristics for TextService and SocialGraphService, two of the Social Network's tiers, which are representative of the other tiers of the service. TextService manages the text users add to composed posts, and SocialGraphService manages follow relationships between users. We do not show each tier due to space constraints, but have validated that the results are similar for them. All applications are generated using profiling data under medium load; \textit{Ditto has not profiled any other load}. We increase the load until the single-tier application or bottleneck tier in the microservice topology saturates in one or more resources (e.g., disk I/O for MongoDB and CPU for the other applications). Since we use a close-loop workload generator for MongoDB and Redis, which only allows one outstanding request per connection, the latency does not increase significantly at high load. While the end-to-end latency of Social Network increases at high load, the latency of TextService and SocialGraphService only increases slightly, since they are not bottleneck tiers.




The upper three rows show IPC, branch misprediction, L1i, L1d, L2, LLC miss rates, and network and disk I/O bandwidth under low, medium, and high load, with average errors across all applications being 4.1\%, 9.9\%, 7.1\%, 5.1\%, 6.9\%, 12.1\%, 0.1\%, 0.1\%, respectively. This indicates that Ditto accurately clones the overall hardware performance metrics. Memcached and NGINX have low IPC under low load because of high branch misprediction, and L1i and L2 misses, while SocialGraphService has high IPC due to fewer LLC misses. At high load, Memcached and Redis have similar metrics to medium load, however the other four applications exhibit different degrees of L2, LLC misses variation. The results illustrate that applications can have very different characteristics under different loads, which are accurately captured by Ditto in their synthetic counterparts. The network and disk bandwidth also conform to the original by faithfully reproducing the system calls. We only show disk bandwidth for MongoDB since other services do not involve disk I/O. The bottom line plot shows the average, 95th, and 99th percentile latencies, which also match the originals, with the p99 diverging at high load, due to the queueing behavior in the network stack at saturation. 

Fig.~\ref{fig:ditto-e2e_latency} shows the end-to-end latency of original and synthetic Social Network when every individual microservice is replaced with a synthetic one. 

\subsubsection{Validation on Varying Platforms}
\label{sec:ditto-validation_varying_platforms}
We also validate the CPU, network and disk metrics and service latency as we vary the hosted platforms. Each application is profiled only on Platform A, and validated on Platforms A, B and C. Figure~\ref{fig:ditto-cpu_platform} shows that the synthetic benchmarks react to platform changes in a similar way to the original applications. More specifically, all six applications have different degrees of L2 cache miss increases on Platforms B and C, due to their smaller L2 cache sizes. Applications running on Platform B, which is an older CPU generation, have consistently lower IPC. When running all microservices of the Social Network on the small-scale Platform C server, the high degree of interference results in high LLC miss rates for TextService and SocialGraphService, both original and synthetic. 
Network and disk I/O bandwidths are identical across platforms, since the amount of data transferred is independent of the platform. The line plots at the bottom show the latency on the three platforms, where the synthetic always matches the original. All applications experience the highest latency on Platform B because it has the lowest IPC. The latency of MongoDB is significantly lower on Platform A because it benefits from the low random access latency of SSDs. In general, the fact that the synthetic applications react to platform changes the same way as the original, without reprofiling, shows that Ditto accurately captures critical, platform-independent features that impact performance. 





\subsection{CPU Top-down Analysis}
\begin{figure}
    \centering
    \includegraphics[width=0.46\textwidth]{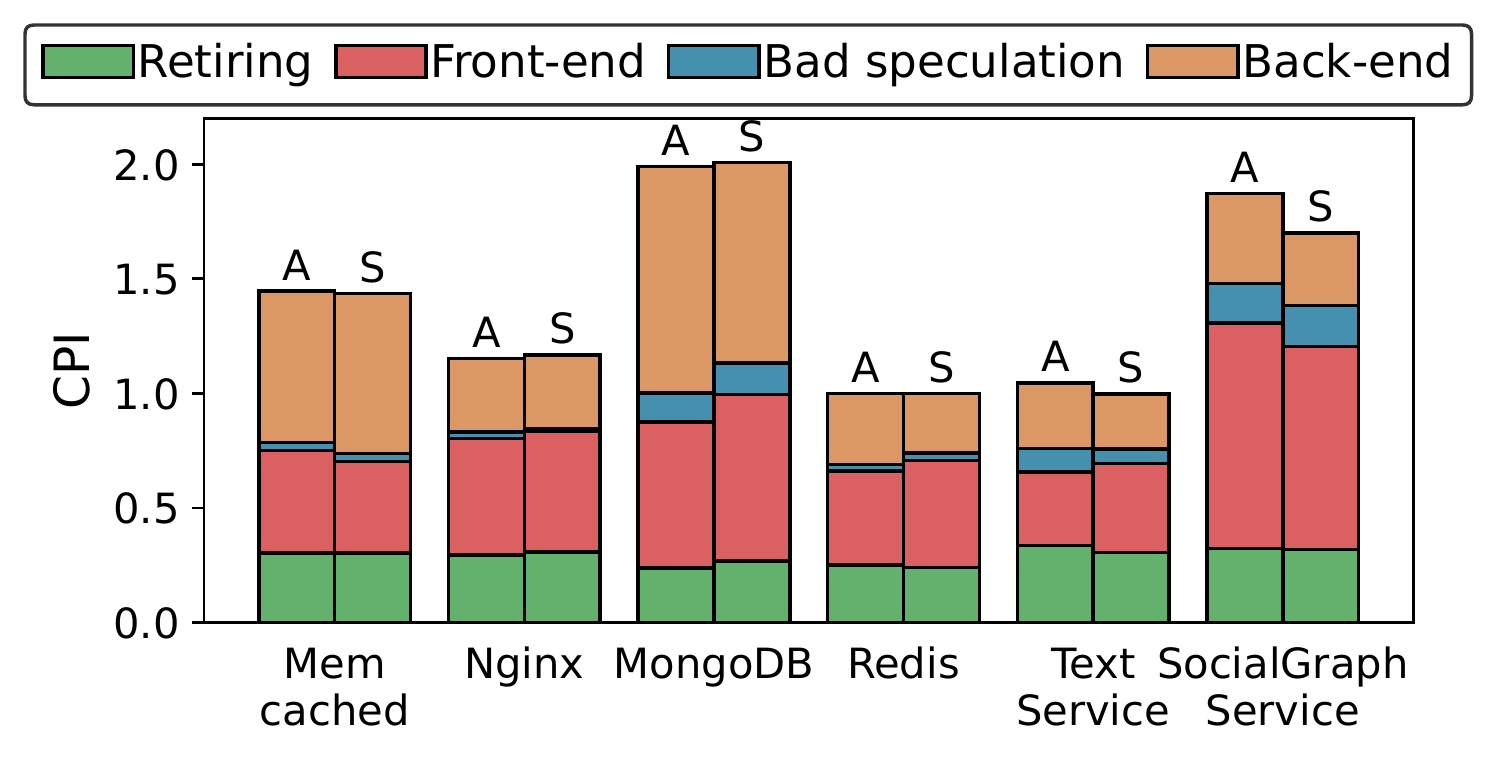}
    \caption{Cycles breakdown. (A: actual, S: synthetic)}
    \label{fig:ditto-applications_topdown}
    \vspace{-0.2in}
\end{figure}

Figure~\ref{fig:ditto-applications_topdown} shows the cycles per instruction (CPI) top-down analysis of the original and synthetic applications. Ditto accurately captures the cycle breakdown of the original applications. Many prior studies have showed that cloud services diverge from traditional scientific CPU benchmarks like SPEC CPU by having significant fractions of front-end stalls, due to large code footprints and frequent context switches between user and kernel mode~\cite{utune,gan:asplos:2019:microservices,reddi15}. Our synthetic benchmarks show similar bottlenecks to the original applications, and can be used as proxies for microarchitectural optimizations. 

\begin{figure}
    \centering
    \includegraphics[width=0.43\textwidth]{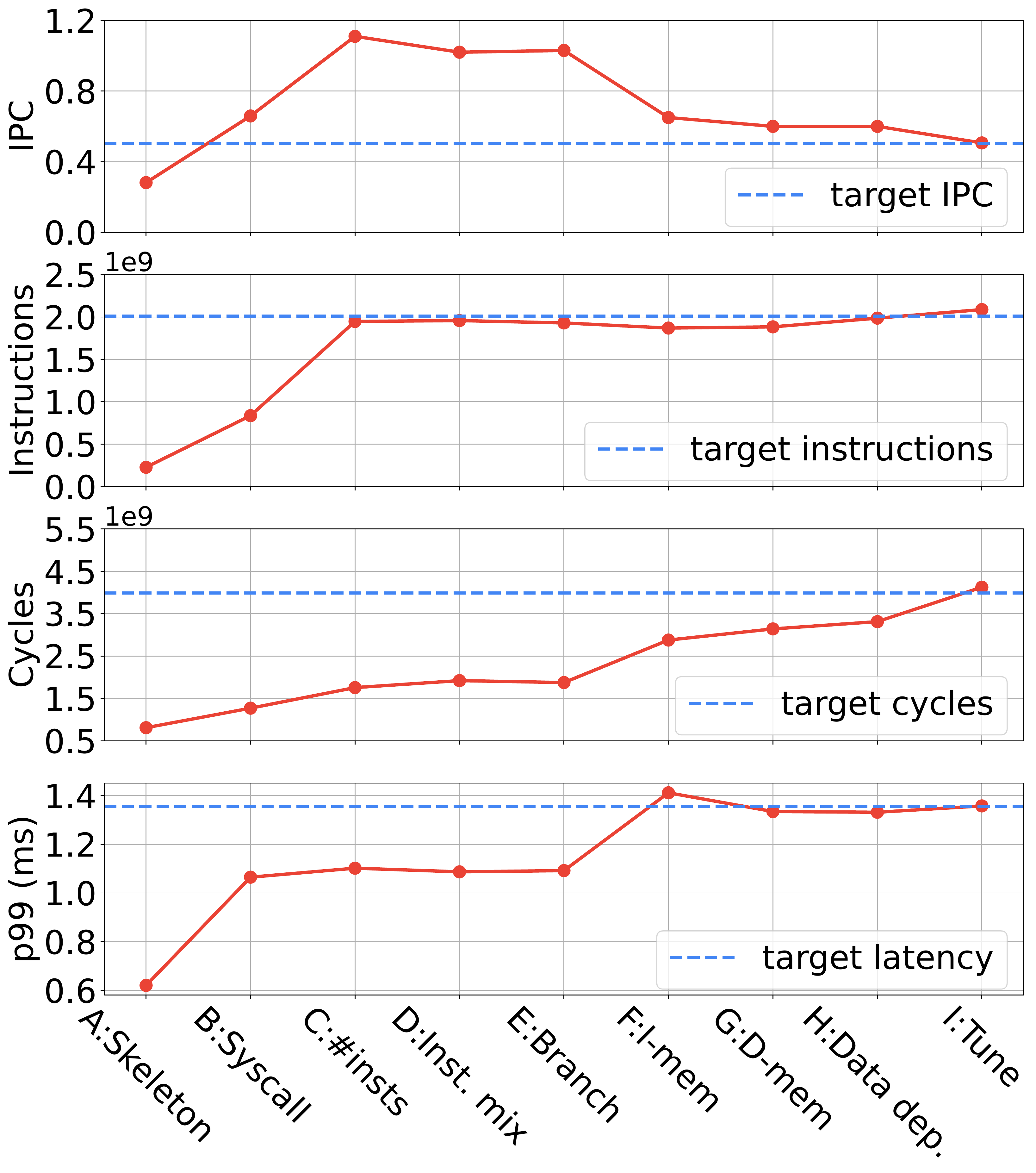}
    \vspace{-0.1in}
    \caption{Evolution of IPC, instructions, cycles, and p99 latency for MongoDB as we add sophistication to Ditto.
    \label{fig:ditto-mongod_incre}}
    \vspace{-0.3in}
\end{figure}

\subsection{Decomposition of Ditto's Accuracy}
In Fig.~\ref{fig:ditto-mongod_incre}, we use MongoDB as an example to show Ditto's accuracy, as the framework incorporates more information. We start with a version of Ditto that only generates the thread model and network interfaces skeleton, but an empty request handling body. From A to B, we inject the system calls with arguments drawn from the distribution of the original application, which increases the kernel-level instructions and disk I/Os. In C, we add user-level instructions (\texttt{add rax, rax}) to match the total instruction count, but not their specific mix. From C to D, user-level instructions are generated based on the profiled mix. We assume the highest branch taken/transition rate, strongest data dependencies, and all memory operations accessing the smallest working sets. We observe an IPC decrease from 1.11 to 1.02 due to memory instructions incurring additional cycles in the backend. From D to E, we clone the branch behaviors following the profiled branch taken and transition rates. The branch misprediction rate drops from 1.95$\%$ to 1.47$\%$ but has a negligible impact on IPC. In step F, we synthesize the instruction memory accesses, which causes more i-cache misses (from 1.3\% to 7.3\%) and branch mispredictions (from 1.47$\%$ to 4.56$\%$, as discussed in Sec.~\ref{sec:ditto-branch_prediction}), and significantly lowers the IPC. From F to G, we synthesize the data memory access pattern by accessing different sizes of private and shared working sets. The IPC further decreases as the L1d miss rate rises from 17$\%$ to 24$\%$. In H, we mimic data dependencies by reassigning registers for each instruction, which clones the ILP and MLP characteristics and slightly lowers the IPC. From H to I, we perform the fine tuning, which calibrates instruction and data access patterns, lowers the IPC from 0.6 to 0.51, and further improves accuracy. This shows that, even if not every aspect in Ditto is equally important, they are all required to accurately clone complex cloud services. 




\begin{figure}
    \centering
    \includegraphics[width=0.40\textwidth]{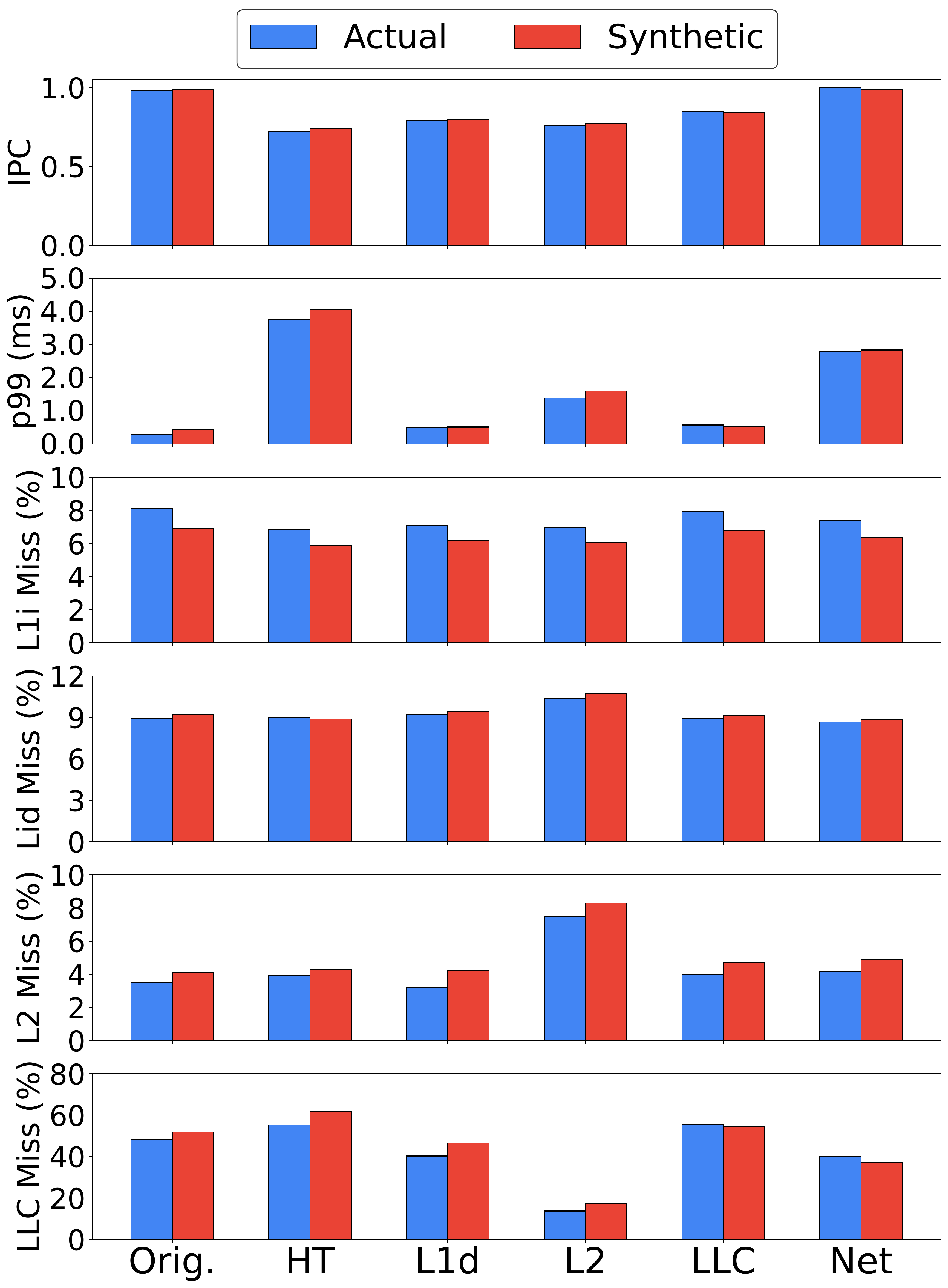}
    \vspace{-0.05in}
    \caption{Interference impact on NGINX. }
    \label{fig:ditto-evaluation_interference}
    \vspace{-0.17in}
\end{figure}

\subsection{Case Study: Interference Analysis}
\label{sec:ditto-evaluation_interference}
Figure~\ref{fig:ditto-evaluation_interference} shows that synthetic applications react to resource interference in a similar way to their original counterparts, even though we only profile the original application in isolation. We show the analysis on NGINX, but the results are similar for other services. We use a set of stress benchmarks to generate interference in different resources. We use stress-ng \cite{stress-ng} to generate hyperthreading (HT), L1d, and L2 interference by co-locating the applications and microbenchmarks on different logical cores of the same physical core. The synthetic application captures the IPC and latency degradation caused by memory contention. When generating L2 interference, besides the L2 miss rate increase, the synthetic workload also captures the LLC miss rate change in the original service due to an increase in the LLC accesses with constant misses. 

We also use iBench~\cite{delimitrou13b} to generate LLC interference on the shared socket, and the result shows the synthetic application captures the IPC drop in the original service. Finally, we use iperf3~\cite{iperf3} to compete with the service for network bandwidth, and the latency of synthetic application successfully matches the original service.




\subsection{Case Study: CPU Core and Frequency Scaling}

\begin{figure}
    \centering
    \includegraphics[width=0.475\textwidth]{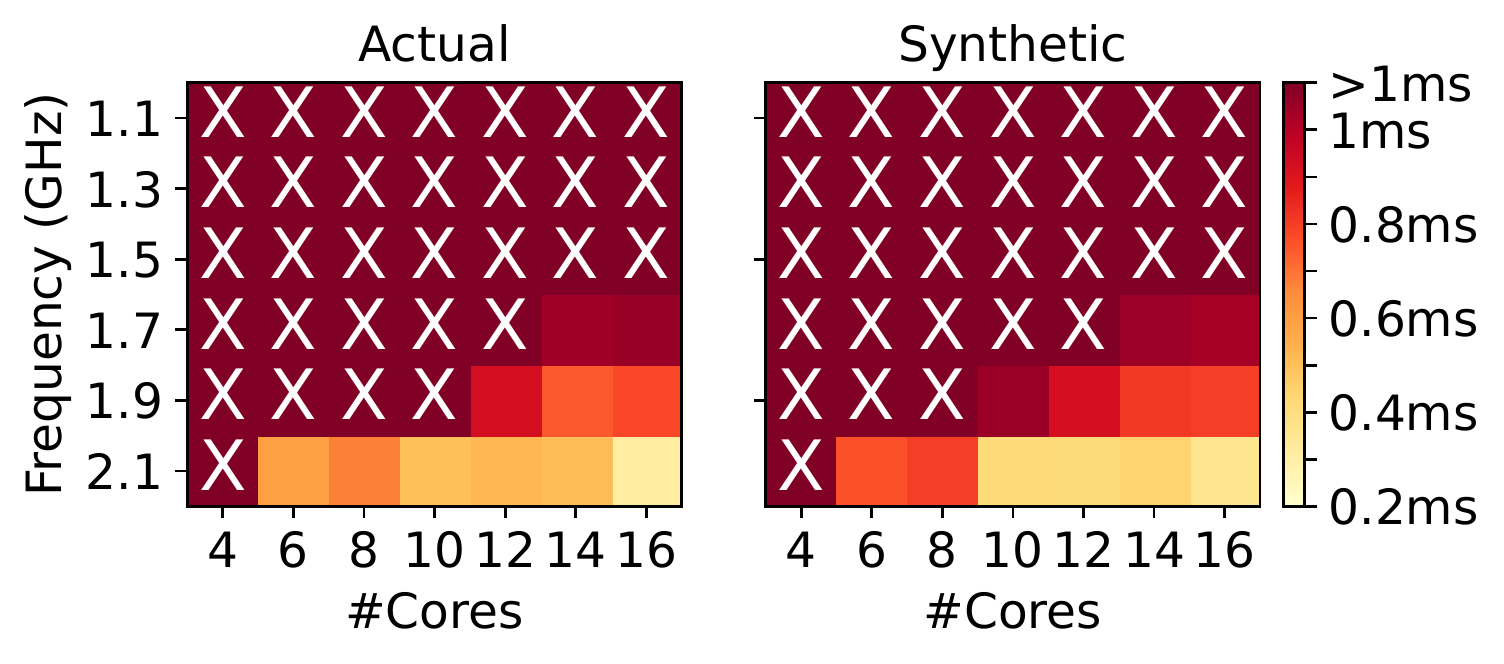}
    \caption{99th percentile latency of actual and synthetic Memcached under varying CPU frequency core count. }
    \label{fig:ditto-memcached_heat}
\end{figure}

Fig.~\ref{fig:ditto-memcached_heat} shows using Ditto to evaluate power management in Memcached with CPU core and frequency scaling. Each cell represents the p99 latency under a given number of cores and frequency. We set the QoS as 1ms and cells with marks mean that QoS cannot be satisfied for that configuration. Memcached cannot meet the QoS at low frequency even with the maximum number of cores, which prohibits aggressive power management. Synthetic Memcached accurately captures the latency variation of Memcached under different settings. This similarity indicates that cloud providers can use synthetic applications to determine whether power management is beneficial for a service, without needing access its source code. 

\section{Conclusion}
\label{sec:ditto-conclusion}

We presented Ditto, an accurate cloning framework for end-to-end monolithic services and microservices. Ditto captures the activity of an application across the system stack, including kernel and network events, and accurately reproduces its characteristics, decoupling representative cloud studies from access to production code. 


\balance

\bibliographystyle{ACM-Reference-Format}
\bibliography{refs}

\end{sloppypar}
\end{document}